\documentclass[fleqn,usenatbib]{mnras}

\usepackage{newtxtext,newtxmath}

\usepackage[T1]{fontenc}

\DeclareRobustCommand{\VAN}[3]{#2}
\let\VANthebibliography\thebibliography
\def\thebibliography{\DeclareRobustCommand{\VAN}[3]{##3}\VANthebibliography}

\usepackage{graphicx}	
\usepackage{amsmath}	
\usepackage{bm} 
\usepackage{ulem}
\usepackage{balance}
\usepackage{enumitem}
\usepackage{multirow}
\usepackage[usenames,dvipsnames]{xcolor}
\usepackage{pifont}

\newcommand{\cmark}{\ding{51}}%
\newcommand{\xmark}{\ding{55}}%

\def\be{\begin{equation}}
\def\ee{\end{equation}}
\def\bea{\begin{eqnarray}}
\def\eea{\end{eqnarray}}

\usepackage{geometry}
\hypersetup{
    colorlinks = true,
    citecolor = {MidnightBlue},
    linkcolor = {BrickRed},
    urlcolor = {BrickRed}
}

\author[M.~Irfan \& P.~Bull]{
Melis O. Irfan,$^{1,2}$\thanks{E-mail: mirfan@myuwc.ac.za}
Philip Bull$^{2,1}$
\\
$^{1}$Department of Physics and Astronomy, University of Western Cape, Cape Town 7535, South Africa \\
$^{2}$Astronomy Unit, Queen Mary University of London, Mile End Road, London E1 4NS, United Kingdom
}

\date{Accepted XXX. Received YYY; in original form ZZZ}

\pubyear{2021}

\begin{document}
\label{firstpage}
\pagerange{\pageref{firstpage}--\pageref{lastpage}}

\title[Cleaning foregrounds from single-dish 21cm intensity maps with KPCA]{Cleaning foregrounds from single-dish 21cm intensity maps with Kernel Principal Component Analysis}
\maketitle

\begin{abstract}
The high dynamic range between contaminating foreground emission and the fluctuating 21cm brightness temperature field is one of the most problematic characteristics of 21cm intensity mapping data. While these components would ordinarily have distinctive frequency spectra, making it relatively easy to separate them, instrumental effects and calibration errors further complicate matters by modulating and mixing them together. A popular class of foreground cleaning method are unsupervised techniques related to Principal Component Analysis (PCA), which exploit the different shapes and amplitudes of each component's contribution to the covariance of the data in order to segregate the signals. These methods have been shown to be effective at removing foregrounds, while also unavoidably filtering out some of the 21cm signal too. In this paper we examine, for the first time in the context of 21cm intensity mapping, a generalised method called Kernel PCA, which instead operates on the covariance of non-linear transformations of the data. This allows more flexible functional bases to be constructed, in principle allowing a cleaner separation between foregrounds and the 21cm signal to be found. We show that Kernel PCA is effective when applied to simulated single-dish (auto-correlation) 21cm data under a variety of assumptions about foregrounds models, instrumental effects etc. It presents a different set of behaviours to PCA, e.g. in terms of sensitivity to the data resolution and smoothing scale, outperforming it on intermediate to large scales in most scenarios.
\end{abstract}




\section{Introduction}
\label{sec:intro}

Forthcoming 21cm intensity mapping surveys promise to survey extremely large volumes of the Universe over a broad span of redshifts with a comparatively high survey speed. Modern radio telescopes have sufficient raw sensitivity to enable this, but a firm detection of the cosmological 21cm signal remains elusive -- while definitive detections have been made in cross-correlation with optical galaxy surveys \citep{Chang:2010jp, Wolz:2015lwa, Anderson:2017ert, Wolz:2021ofa}, it has not yet been possible to obtain a reliable measurement of the 21cm auto-power spectrum, or the baryon acoustic oscillation feature that is their principal scientific target \citep{Chang:2007xk, Bull:2014rha}. The main reason for this boils down to the very large dynamic range of the data, with foreground contamination around $3-4$ orders of magnitude larger than the cosmological signal.

While foregrounds are in principle distinguishable from the rapidly spatially and spectrally-varying 21cm fluctuations by their smooth frequency spectra, approaching a power-law behaviour characteristic of Galactic and extragalactic synchrotron emission. This is significantly complicated by instrumental effects and calibration errors however, which modulate and scatter the spectrally-smooth foregrounds into other regions of Fourier space, thus contaminating what would otherwise be `clean' signal-dominated modes. The detailed behaviours of these effects can depend heavily on the particular instrument in question, but some reasonably generic issues can be identified. For example, outside the main lobe of the primary beam pattern of the telescope, near zero-crossings (nulls) tend to arise, with their exact angular distance from the beam centre depending on frequency. Bright sources close to the null at one frequency may therefore pass through it at another frequency, strongly modulating the otherwise smooth spectrum of the source \citep{ghosh}. Similarly, the leakage of polarised foreground emission into the unpolarised channel due to beam and receiver imperfections can also generate additional spectral structure; polarised emission is inherently more spectrally complex due to Faraday rotation and other line-of-sight effects, and this can be exacerbated by the additional spectral structure of the beam imperfections themselves \citep{shaw, alonsoPol}.

The most common approach to dealing with the foreground contamination issue is to apply a so-called `blind' foreground subtraction algorithm to the data \citep{Wolz:2013wna, alonso, olivari, isa, steve}. These methods do not require a model of the data to be specified; instead, they perform transforms of various kinds in order to try and segregate the foreground information from the cosmological 21cm signal (and noise, which is also rapidly fluctuating). A broad class of such algorithms is based on Principal Component Analysis (PCA), which calculates the empirical frequency-frequency covariance matrix of the data (i.e. by averaging over angular pixels), and then performs an eigendecomposition to arrive at a set of distinct modes that can, in principle, be segregated according to their signal-to-noise ratio. Since the foreground emission is vastly brighter than the 21cm signal, this should allow the foregrounds to be filtered out simply by discarding the modes with the largest eigenvalues (i.e. the biggest SNR). A number of other unsupervised foreground removal methods use a similar filtering approach to PCA while using alternative means to arrive at a set of appropriate basis functions. For example, Independent Component Analysis \citep[e.g. ICA;][]{Wolz:2013wna}, uses a cost function that maximises the non-Gaussianity of the recovered modes in addition to a eigenbasis decomposition. In principle, this allows the non-Gaussian-distributed foregrounds to be more cleanly separated from the approximately Gaussian 21cm signal and noise. Conversely, Gaussian Progress Regression \citep[GPR;][]{mertens, Ghosh:2020fpy, kern, paula}, which is a non-parametric Bayesian method, chooses to encapsulate the 21cm contaminants as a combination of Gaussian variables so as to capture stochastic contaminates, such as calibration errors and contributions from the ionosphere. Other methods perform the component separation in different bases in which a cleaner separation of signal and foreground is possible such as GMCA \citep{gmca, chapman, isa}, which uses sparsity to enforce the separation of the different components, and GNILC \citep{firstGNILC, olivari}, which performs the eigendecomposition in the needlet domain. 

These methods are largely successful at removing the bulk of the foreground emission, but at a cost. The high-SNR foreground modes tend to have a similar spectral structure to the longest-wavelength radial Fourier modes (i.e. those at lowest $k_\parallel$) in the 21cm signal. The signal modes are therefore filtered out as well, leading to suppression or even total loss of the cosmological 21cm signal on some range of scales \citep{Chang:2010jp, alonso}. Similar over-fitting also affects the noise. Because instrumental and other systematic effects scatter the foregrounds across a broader region of Fourier space, the foreground filters must use an increasing number of modes or more complex basis functions to be able to remove the foregrounds well, leading to progressively larger signal loss that affects an even wider range of scales. In realistic applications, some methods may require tens of modes to be subtracted from data with only a couple of hundred frequency channels, resulting in substantial signal loss even at relatively high $k$ values \citep[e.g.][]{Wolz:2015lwa, Wolz:2021ofa}. This is often corrected for by performing simulated signal injection on the data to identify which (signal) Fourier modes are being suppressed and to what level. This information can then be used to construct a `transfer function' that corrects the measured power spectrum \citep{tfunc1, tfunc2, Wolz:2021ofa}. The effectiveness of this procedure depends to some extent on the realism of the injected simulations and how well they reproduce instrumental filtering effects.

For cross-correlation analyses, somewhat less aggressive foreground filtering can be applied. Since the foregrounds are in principle uncorrelated with the non-21cm survey, e.g. an optical galaxy survey, they will tend to average out in the cross-correlation, and so a moderate foreground residual can be left in the 21cm data, reducing the amount of signal loss due to filtering \citep{21opt, steve19}. Leaving too large a residual will impact the noise properties of the cross-correlation however, which goes approximately like the product of the square roots of the noise variances of the individual surveys \citep[e.g.][]{Villaescusa-Navarro:2014rra}, where in this case the foreground residual would be included as part of the 21cm survey's noise variance. 

In this paper, we study an extension of PCA filtering to a broader class of algorithms called Kernel PCA \citep[KPCA;][]{scholkopf1997kernel}. These allow arbitrary sets of non-linear transformations to be applied to the data, making it possible to construct more complicated decompositions of the frequency-frequency covariance and therefore -- in principle -- provide a cleaner separation of the foregrounds and 21cm signal. KPCA methods are commonly used in other domains for segmentation and classification, where `decision boundaries' between different sub-populations of the data are not well-described by linear functions of the data dimensions. The choice of kernel function determines which kinds of nonlinear transformation can be applied, and once a few hyperparameters have been specified, no further human intervention is needed, making this a class of `blind' methods also.

We consider whether KPCA can provide a better separation between the 21cm and foregrounds in the presence of instrumental effects than linear PCA-type methods. We use simulated data for a MeerKAT-like 21cm survey that has been subjected to semi-realistic beam convolutions, and which includes realistic data-driven foreground models with complex spectral structure and masking effects, comparing our results with the well-known PCA and ICA methods. We also consider if improvements can be gained by smoothing the data, at each, frequency to a common resolution. We present our results in terms of the recovered (residual) power spectrum of the data, and also the recovered cross-power spectrum with a fictitious galaxy survey at the same redshifts.

This paper is structured as follows. In Sect.~\ref{sec:methods} we review the PCA and Kernel PCA methods, and the choices that can be made when configuring Kernel PCA, such as the choice of kernel and a possible pre-cleaning step. In Sect.~\ref{sec:sims}, we describe the properties of the simulated 21cm data cubes that we use in this study, including our models for the matter, 21cm, and galaxy density fields, the instrumental beam, and a selection of foreground models. In Sect.~\ref{sec:results}, we present our results, comparing the Kernel PCA method with PCA and ICA in a number of different scenarios. We conclude in Sect.~\ref{sec:conclusions}. Throughout this paper, we assume a $\Lambda$CDM cosmology with parameters $\Omega_{\rm cdm} = 0.25$, $\Omega_b = 0.05$,
$h = 0.7$, $n_s = 0.95$, and $\sigma_8 = 0.8$.

\section{Foreground removal with Kernel PCA}
\label{sec:methods}

In this section, we introduce Kernel PCA as a foreground removal technique for 21cm intensity mapping data. We first review the use of standard PCA as a foreground filter, before describing how Kernel PCA works and demonstrating the effect of different choices of kernel.

\subsection{Principal Component Analysis}

PCA is an unsupervised method that can be used to construct a suitable basis of frequency-dependent functions that can then be used to project the foregrounds out from the data. It works by finding the eigenbasis decomposition of the empirical frequency-frequency covariance matrix of the (mean-subtracted) data, i.e. averaging over the available angular pixels,
\be
C(\nu, \nu^\prime) = \frac{1}{N_{\rm pix}} \sum_j^{N_{\rm pix}} \delta T_j(\nu) \delta T_j(\nu^\prime).
\ee
The modes with the largest eigenvalues correspond to the largest sources of variance in the data. Since the foregrounds are orders of magnitude brighter than the 21cm signal, they are expected to be the dominant source of variance and therefore mostly confined to these large-eigenvalue modes. We can then subtract the projection of the data in each pixel onto these modes, leaving behind a lower-variance residual that should in principle contain only 21cm signal plus noise. Because the foregrounds tend to have smooth (power law-like) frequency spectra, these modes will also tend to be smooth functions of frequency.

In reality, there is not a perfect split between the foregrounds and the 21cm signal; while substantially different in terms of their variance and spectral structure, they are not completely orthogonal. Some of the smoother radial modes in the 21cm signal, corresponding to low values of $k_\parallel$ (and also low $k_\parallel$), can therefore also be projected out with the foreground modes, leading to loss of signal power. Depending on the number of PCA modes that are subtracted, this can substantially suppress the recovered 21cm power spectrum over a range of scales. Typically, this signal loss is then corrected for by using the effect of the foreground filter on a known signal injected into the data to determine a signal loss transfer function, which can then be divided out from the measured power spectrum.

A further complication is that other spurious spectral structure can also be present in the data. The instrumental beam can modulate the sky signal in a way that mixes components and induces additional spectral structure for example \citep{siya}, while polarisation leakage can add non-smooth polarised foregrounds into temperature-only data. This requires more PCA modes to be subtracted, thus increasing the amount of signal loss. In the next section, we introduce Kernel PCA, a non-linear method that is capable of constructing more complicated ways of separating components.

\subsection{Kernel PCA}

Kernel PCA \citep{scholkopf1997kernel} is an extension of PCA that operates on a non-linear transformation of the original data vectors. The transformed space (called the {\it feature space}) can have an arbitrarily higher dimension than the original data. In general, this makes it easier to find an effective segregation between different components of the signal, which can be achieved simply by using standard linear PCA in the higher-dimensional feature space.

As an example, consider a 21cm data cube that we split up into a set of data vectors, one frequency spectrum (of dimension $N_{\nu}$) per pixel on the sky, which we denote $\vec{x}_j$ for pixel $j$. A non-linear transformation is then applied, $\vec{y}_j = \Phi(\vec{x}_j)$, resulting in a transformed vector $\vec{y}_j$ of dimension $N_{\rm feature}$. Because the transformation $\Phi$ is non-linear, the feature space now has dimensions that are various non-linear combinations of the original dimensions of the data; for example, for a spectrum with $N_\nu = 2$ and original dimensions $\{ \hat{x}_{0}, \hat{x}_{1} \}$, one could form a feature space with dimensions $\{ \hat{x}_{0},\, \hat{x}_{1},\, \hat{x}_{0} \hat{x}_{1},\, \hat{x}_{0}^2,\, \hat{x}_{1}^2,\, \dots\}$, depending on the choice of $\Phi$.\footnote{To clarify our notation, we denote basis vectors as $\hat{x}_n$ and their coefficients as $x_n$, for each $n = 0 \dots N_\nu-1$, so that $\vec{x}_j = x_{0j} \hat{x}_0 + x_{1j} \hat{x}_1 = (x_{0j}, x_{1j})$.} Linear combinations of these dimensions can then be found that decompose the feature-space covariance matrix,
\be
\tilde{C} = \frac{1}{N_{\rm pix}} \sum_j^{N_{\rm pix}} \vec{y}_j \vec{y}_j^{\rm T} = \frac{1}{N_{\rm pix}} \sum_j^{N_{\rm pix}} \Phi(\vec{x}_j) \Phi^{\rm T}(\vec{x}_j),
\ee
into principal components. $\tilde{C}$ has dimensions $N_{\rm feature} \times N_{\rm feature}$, which can be very large. When these components are projected down onto the original lower-dimensional space of the data, the resulting functions can have more complicated functional forms than the simple linear forms that standard PCA is able to use, making this method more flexible in its ability to separate different components of a signal. To take a simple example, consider a set of datapoints that trace out a circle in the 2D space $\{\hat{x}_{0}, \hat{x}_{1}\}$. Linear PCA would not be able to provide a good description of this distribution, as it can only draw a straight line through it (i.e. a linear function of the data dimensions). The Kernel PCA method with a non-linear transformation $\Phi$ that generates dimensions $\hat{x}_{0}^2$ and $\hat{x}_{1}^2$ (amongst others) would easily be able to trace out a circle however.

The non-linear transformation $\Phi$ can result in an arbitrarily large feature space, making calculations unwieldy; explicit computation of the covariance matrix may be intractable for example. Fortunately, it is not necessary to work directly in the feature space. We can instead use the so-called {\it kernel trick} to rewrite the inner product of the transformed vectors as
\be
k(\vec{x}, \vec{x}^\prime) = \Phi(\vec{x}) \cdot \Phi(\vec{x}^\prime),
\ee
where $k$ is the {\it kernel}.
The following matrix can then be formed by using the kernel to calculate the inner product of all pairs of transformed data vectors,
\be
K_{ij} = k(\vec{x}_i, \vec{x}_j) = \Phi(\vec{x}_i) \cdot \Phi(\vec{x}_j),
\ee
which has shape $N_{\rm pix} \times N_{\rm pix}$. (The transformed data vectors are typically also mean-centred before this step, i.e. $\Phi(\vec{x}_j) \to \Phi(\vec{x}_j) - (1/N_{\rm pix}) \sum_i \Phi(\vec{x}_i)$.) 
Only this object is needed to perform a PCA decomposition of the feature-space covariance matrix, and not the covariance matrix itself.\footnote{We need not even define an explicit form for $\Phi$; by choosing a form for the kernel $k$, a particular $\Phi$ is implicitly defined.} The resulting eigenvectors in the feature space do not need to be evaluated explicitly either; instead, we can project an arbitrary vector in the data space onto the higher-dimensional principal components, again using the matrix $K_{ij}$ to evaluate the necessary inner products. This makes working with even infinite-dimensional non-linear transformations tractable -- any calculations involving inner products in the feature space can be evaluated simply by evaluating the kernel function for pairs of vectors in the much lower-dimensional data space, rather than having to go into the feature space and explicitly sum the individual components of the inner product for $N_{\rm feature}$ dimensions. For more details, and simple proofs of these statements, see \citet{scholkopf1997kernel}.

To implement a foreground filter with Kernel PCA, we can then perform a similar procedure to standard PCA and simply subtract off templates constructed from the first few eigenmodes of the feature-space covariance from each pixel. In this case, the modes used are now the {\it projections} of the feature-space eigenmodes onto the data. While the templates themselves are constructed from non-linear functions of the dimensions of the data, the subtraction step itself is a linear operation.

Finally, we recall that the size of the inner product matrix $K_{ij}$ ($N_{\rm pix} \times N_{\rm pix}$) can be quite large for a typical dataset, and so operations on this matrix tend to dominate the computational cost of the Kernel PCA algorithm. It should therefore be noted that Kernel PCA can have a substantially higher computational cost than standard PCA on the same dataset.

\subsection{Choice of kernel}
Different choices of kernel generate different feature spaces, which may be more or less suited to segregating particular mixtures of signal depending on the functional forms that they allow. The kernels themselves also have hyperparameters that can be tuned to produce better results, e.g. by promoting more or less smoothness of the functional forms, fixing a typical length-scale and so on. Common choices of kernel include polynomials, radial basis functions (Gaussians), and sigmoid ($\tanh$) functions, amongst others.

In this paper, we compare these three common choices of kernel to make an empirical determination of suitability to the 21cm foreground removal problem. We do not seek to find optimal choices of kernel or hyperparameters, leaving this to future work.

We use the following kernels, as implemented in the {\tt decomposition.KernelPCA} function of {\tt scikit-learn} \citep{scikit-learn}:
\begin{align}
k(\vec{p}, \vec{q}) = & ~~(\gamma \vec{p}^{\rm T} \vec{q} + c)^d & {\rm (Polynomial)} \\
 & \exp \left ( -\gamma ||\vec{p} - \vec{q}||^2\right ) & {\rm (Gaussian/RBF)} \\
 & \tanh(\gamma \vec{p}^{\rm T} \vec{q} + c) & {\rm (Sigmoid)}.
\end{align}
The hyperparameters $\gamma$, $c$, and $d$ control the scale of the kernel, an offset, and the maximum order of the polynomial basis respectively. The kernel scale ($\gamma$) controls the curve of the decision boundary, which is the hypersurface that separates out the different components within feature space \citep{gamma}. The way that each of these parameters changes the detailed behaviour of the kernel is kernel-dependent, and the best choice of kernel for a particular problem is typically selected through some combination of trial and error and hyperparameter optimisation. \citet{pickingKernel} present an unsupervised method of selecting the optimum kernel for use by first transforming the data into feature matrices, using numerous different kernels, and then finding the kernel combination that maximises the Frobenius norm of the column-wise and row-wise elements of the feature matrices. Although the data themselves determine the optimum kernel and hyperparameter choices, these choices are not physically intuitive. We do not explore this issue further here.
\subsection{Pre-cleaning}
\label{sec:sims}
Even the most basic foreground removal methods are broadly successful at reconstructing the true 21cm signal, producing residuals of order the size of the cosmological signal or less despite the foregrounds themselves being several orders of magnitude larger. The bigger challenge for foreground removal methods are subtler features in the data with more complex spectral signatures, such as those generated by the interaction between foreground emission and frequency-dependent beam sidelobes.

We are primarily interested in how Kernel PCA performs in the presence of these features. As such, we focus our analysis on the data {\it after} the bulk of the foreground structure has already been removed, or``pre-cleaned'', using a rough initial pass of another foreground cleaning method. Ideally, the pre-cleaning technique would reduce the variance of the foreground fluctuations from four orders of magnitude larger than the 21\,cm signal fluctuations to around the same order of magnitude (e.g. perhaps a few times larger), while leaving the 21\,cm signal as intact as possible at this stage, i.e. minimising the signal loss. The desired behaviour can be verified by eye, to check that the pre-cleaned maps display temperature fluctuations larger than a few times the expected level of the 21cm signal for example. We use a simple linear PCA method that removes only two eigenmodes to do the pre-cleaning, but any technique capable of reducing the foreground variance by several orders of magnitude without over-fitting would be suitable.

Unlike some other methods, we note that this pre-cleaning step appears to be necessary for KPCA; without it, substantially larger residuals are obtained, even when the number of subtracted modes is increased and different kernels and hyperparameters are used. While we do not have a detailed explanation for this behaviour, we anticipate that it is caused at least in part by the sensitivity of non-linear decompositions to signals separated by a large dynamic range. Pre-cleaning improves and stabilises the behaviour substantially, suggesting that KPCA's strengths are in separating complicated signals of a similar size rather than in reducing the dynamic range of the data.

\section{Simulations of auto-correlation surveys}
\label{sec:sims}

In this section, we describe a set of simple simulated 21cm auto-correlation survey datacubes that we use to test Kernel PCA. We simulate a 3600 degree square region and cover the frequency range of 950.6 -- 1234.7\,MHz; replicating a cosmological box of dimensions $1.25 \times 1.25 \times 1.25$\,Gpc in the redshift range $z = 0.150 - 0.494$. The simulated 21cm field is based on Gaussian realisations of the matter power spectrum with appropriate linear bias factors and a log-normal transform applied. We also add two different sets of foreground models, frequency-dependent Gaussian instrumental noise, and convolve the fields with a realistic frequency-dependent beam model, based on the specifications of the MeerKAT array. We choose to focus on the MeerKAT experimental set-up due to the recently proposed 21cm intensity mapping survey: MeerKLASS \citep{mklass}, which is currently underway \citep{wang21}. 

\subsection{Neutral hydrogen distribution}
\label{sec:nhd}

On large scales, beyond the non-linear regime, the matter distribution is expected to be approximately Gaussian. Apart from at the lowest redshifts, the beam smoothing for MeerKAT and SKAO in auto-correlation mode is sufficiently large that the density field can only be recovered on large linear scales anyway, and so non-linearities are only a minor concern. As such, we would expect Gaussian density fields to be a good approximation for essentially all of the scenarios that we consider in this paper. This allows us to use a particularly simple method to generate simulated density fields:
\begin{enumerate}
  \item Draw unit Gaussian random numbers in each cell of a cubic grid and perform an FFT.
  \item In Fourier space, multiply the value in each grid cell by the square root of the matter power spectrum, $P(k, z)$.
  \item After multiplying by an appropriate normalisation, perform the inverse FFT to recover a Gaussian random realisation of the matter density field.
\end{enumerate}
We use {\tt CCL}\footnote{\url{https://github.com/LSSTDESC/CCL}} \citep{ccl} to calculate the matter power spectrum, and allow the grid to have different side lengths in each dimension.

Pathologies can arise in the Gaussian realisations when examining smaller scales however. For example, the value of the density contrast, $\delta_m$, can drop below $-1$ in some grid cells if the grid cell size is small enough. One way of fixing this issue is to apply a log-normal transformation to the Gaussian density field \citep{Coles:1991if, Agrawal:2017khv}, which guarantees that the density contrast remains in the physical range, $\delta_m \in [-1, \infty]$. This also provides a simple model of the non-linear density field, yielding a configuration-space matter power spectrum that is accurate to within about 10\% at $k = 0.2\, h/{\rm Mpc}$ \citep{2015MNRAS.452..686C}. In all that follows, we use log-normal transformations to model the matter and 21cm brightness temperature fields, as well as the galaxy number counts discussed in Sect.~\ref{sec:mockgalaxy}.

For the 21cm brightness temperature field, we adopt a simple linear prescription to map the matter density contrast to brightness temperature fluctuations, such that
\be \label{eq:deltatb}
\Delta T_b(\vec{x}, z) = \overline{T}_b(z)\, b_{\rm HI}(z)\, \delta_m(\vec{x}, z).
\ee
The mean brightness temperature, HI bias, and fractional HI density are given by the following fitting formulas, calculated from the model described in \cite{Bull:2014rha}:
\bea
\overline{T}_b(z) &=& 0.055919 + 0.23242\, z - 0.02414\, z^2 \nonumber \\
b_{\rm HI}(z) &=& \frac{b_{{\rm HI}, 0}}{0.677105} \left (0.66655 + 0.17765\, z + 0.050223\, z^2 \right ) \nonumber \\
\Omega_{\rm HI}(z) &=& \frac{\Omega_{{\rm HI},0}}{4.86} \left ( 4.8304 + 3.8856\, z - 0.65119\, z^2 \right ),
\eea
where we set $\Omega_{{\rm HI},0} = 4.86 \times 10^{-4}$ and $b_{{\rm HI}, 0} = 0.677105$. We incorporate redshift-space distortions by shifting each voxel of the log-normal transformed field to a new radial position,
\be
s_\parallel = x_\parallel - \frac{{\rm v}_\parallel + {\rm v}_{\rm NL}}{H(z_c)},
\ee
where $x_\parallel$ is the original radial coordinate of the voxel centre, ${\rm v}_\parallel$ is the velocity in the radial direction (calculated from the matter density field using the linear continuity equation), the non-linear velocity ${\rm v}_{\rm NL}$ in each voxel is drawn from an uncorrelated Gaussian random distribution with standard deviation $\sigma_{\rm NL} = 120\,{\rm km/s}$, and the expansion rate $H(z)$ is evaluated at the central redshift of the simulation box, $z_c$. After applying this shift, the voxels are interpolated back onto a regular grid. The expression in Eq.~\ref{eq:deltatb} can then be updated to its redshift-space equivalent,
\be
\Delta T_b(\vec{s}, z) = \overline{T}_b(z_c)\, \delta_s(\vec{s}, z_c),
\ee
where $\delta_s$ is the biased log-normal transform of the HI density contrast in redshift space.

With the redshift-space brightness temperature distribution in hand, we then project the simulated box into observational coordinates, i.e. redshifts/frequencies and angles. We employ the distant observer approximation, evaluating the relevant cosmological quantities at the central redshift of the box, and ignoring their evolution within the box. Using this approximation, we also neglect curved sky effects, such that each frequency slice occupies the plane perpendicular to the line of sight connecting the observer to the centre of the simulation box. Since the solid angle subtended by the simulation volume ($60^\circ \times 60^\circ$) is quite large and the central redshift of $z_c \approx 0.32$ (corresponding to a comoving distance to the centre of the box of $1.28$~Gpc) is reasonably small, this approximation is expected to be limited in accuracy on the largest scales we consider. We do not expect this to significantly impact our conclusions however.

\subsection{Instrumental beam}

The angular and spectral structure of the observed foregrounds is strongly affected by the behaviour of the instrumental beam that the sky signal is convolved with. At its most basic, the effect of the beam is to introduce a frequency-dependent angular smoothing, of characteristic scale $\theta \sim \lambda / D_{\rm dish}$, where $\lambda$ is the wavelength of observation, and $D_{\rm dish}$ is the diameter of each dish ($\sim 13.5$m for MeerKAT). This has the effect of mixing together the signals from foreground regions that may have different spectral indices etc, resulting in a slightly more complex spectral behaviour while smoothing away any small-scale structure.

The existence of frequency-dependent sidelobes (secondary, lower-level peaks in sensitivity) and nulls (zeros or low-sensitivity regions) in the beam, at angles far from the pointing centre, further complicates the spectral structure by modulating the otherwise smooth spectra of sources that fall within these regions. While the amplitude of the sidelobes is typically much lower than the centre of the beam, they can cover a large solid angle, and so contribute a non-negligible fraction of the total detected emission after integrating over the entire beam pattern. Bright sources that move in and out of sidelobe/null regions as a function of frequency can also contribute substantially to the total signal, with a frequency spectrum that will be strongly modulated. While other, related, sources of foreground chromaticity also arise, such as gain errors and polarisation leakage, we will only include beam chromaticity in this paper.

\begin{figure}
	\includegraphics[width=1.1\columnwidth]{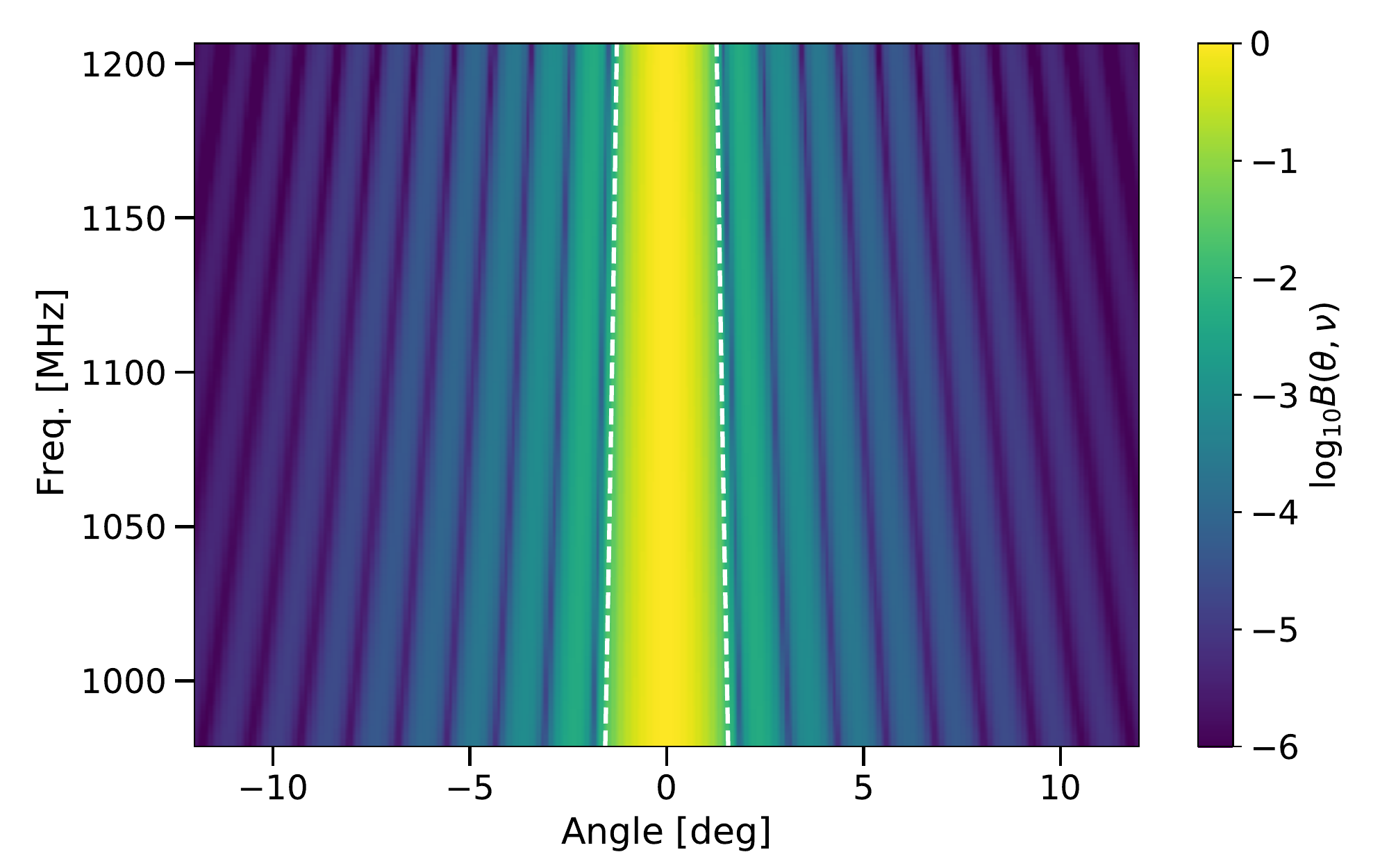}
    \caption{A slice through the beam model for the MeerKAT L-band (Stokes I polarisation) from {\tt katbeam}, as a function of angle and frequency. The white dashed lines show the approximate beamwidth as a function of frequency, $\Delta \theta \approx 1.2 \lambda / D_{\rm dish}$.}
    \label{fig:beam}
\end{figure}

For our beam model, we adopt the L-band, Stokes I polarisation beam pattern from the {\tt katbeam} package\footnote{\url{https://github.com/ska-sa/katbeam}}, which provides a simplified model of the MeerKAT beam with substantial sidelobe structure and a small frequency ripple in the beam width, as described in \cite{asad21}. The beam model is convolved with the simulated sky separately in each frequency channel using a 2D FFT, which preserves the mild asymmetry of the beam while assuming periodicity of the survey area in each frequency channel. We do not attempt to model a scanning strategy for the survey, which would result in the measured intensity in each pixel deriving from a mixture of beam convolutions at different angles, so the beam convolution we have applied is also simplified in this sense. The beam model as a function of frequency is shown in Fig.~\ref{fig:beam}.

\subsection{Foreground models}

\begin{figure*}
	\includegraphics[width=1.8\columnwidth]{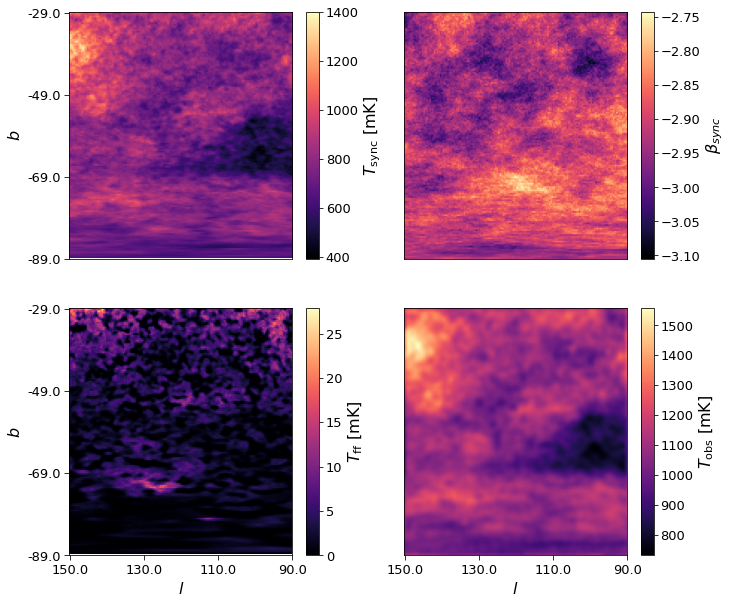} \\
    \caption{The Planck Sky Model (PSM) foregrounds in Galactic coordinates. {(\it Upper left):} Synchrotron emission amplitude at 950.6\,MHz. {(\it Upper right):} Synchrotron spectral index. {(\it Lower left):} Free-free emission amplitude at 950.6\,MHz. {(\it Lower right):} Combined diffuse plus point source contributions at 950.6\,MHz, after convolution with the beam.}
    \label{fig:forepsm}
\end{figure*}

Bright foreground contamination -- largely from diffuse Galactic synchrotron emission and extragalactic radio sources -- is the primary systematic effect in 21cm surveys. While the foreground emission itself is expected to have a smooth frequency spectrum in total intensity, additional spectral structure is imparted by calibration errors and the structure of the instrumental beam. Polarised emission, with a more complex spectral structure due to effects such as Faraday rotation, can also leak into the total intensity channel, further complicating the foreground signal.

We consider two different models of foreground contamination in this work: the Global Sky Model (GSM) and the {\it{Planck}} Sky Model (PSM). These models are constructed in quite different ways, and so by comparing them we can get a fair picture of the performance of various foreground removal methods, without needing to worry that they might be over-optimised for a particular foreground model, e.g. due to exploiting unrealistic structures.

The GSM \citep{gsmref} is a publicly-available algorithm\footnote{\url{https://github.com/telegraphic/pygdsm}} capable of producing all-sky HEALPix maps of diffuse Galactic emission at any frequency between 10\,MHz and 5\,THz. The GSM combines information from 29 different sky surveys, each covering different regions of the sky and frequencies as well as having different resolutions, and uses PCA to pick out the major sky components and interpolate them across the full frequency range and sky area. Below 10 GHz, GSM is capable of producing maps at a resolution of either 5 degrees or 56 arcmins; we use the 56 arcmin simulations.

The PSM, as implemented in the {\it{Planck}} full sky simulations \citep{ffp8}, is also publicly-available in the form of the ``full focal plane'' simulations on the {\it{Planck}} Legacy Archive.\footnote{\url{https://pla.esac.esa.int/\#maps}} While the GSM simulates the total diffuse sky emission present at each frequency, the PSM provides separate component maps for (e.g.) free-free and synchrotron emission. We assemble a set of foreground maps from the PSM using the method detailed in \cite{isa, steve}. The pertinent details of our PSM model are summarised in Table~\ref{table:fgdets}, and the individual components are shown in Fig.~\ref{fig:forepsm}.

\begin{table}
 \begin{tabular}{llll}
 \hline
 {\bf FG component} & {\bf{Value}} & {\bf Reference} \\
 \hline
Free-Free amp. & varies & \citet{ffp8} \\
Free-Free $\beta$ & $-2.10$ & \citet{ffform} \\
Synchrotron amp. & varies & \citet{ffp8} \\
Synchrotron $\beta$ & varies & \citet{mamd}  \\
PS cut-off flux & 0.1\,Jy & \cite{olivari} \\
PS $\beta$ & $-2.70 \pm 0.2$ & \cite{olivari} \\
 \hline
 \end{tabular}
 \caption{Components within our implementation of the PSM foreground model and their parameters.}
 \label{table:fgdets}
\end{table}

\begin{figure}
	\includegraphics[width=0.98\columnwidth]{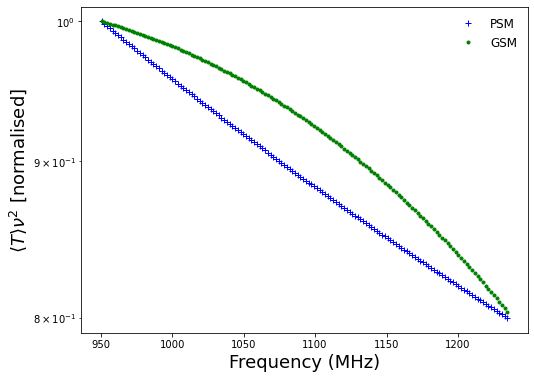} \\
    \caption{The mean emission temperature at each frequency from the GSM and PSM foreground simulations within the high Galactic latitude region investigated in this work; $(l,b) = (120^{\circ}, -59^{\circ})$. The temperature curves are multiplied by frequency squared to emphasis any spectral discrepancies between the two models, and are normalised to unity at the lowest frequency, 950.6\,MHz.}
    \label{fig:gsmpsm}
\end{figure}

\begin{figure}
	\includegraphics[width=0.9\columnwidth]{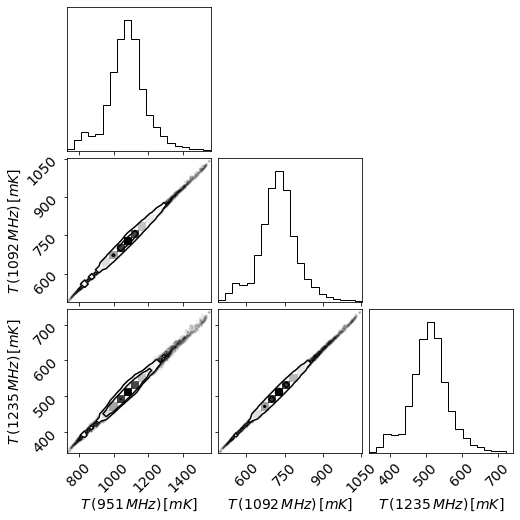} \\
	\includegraphics[width=0.9\columnwidth]{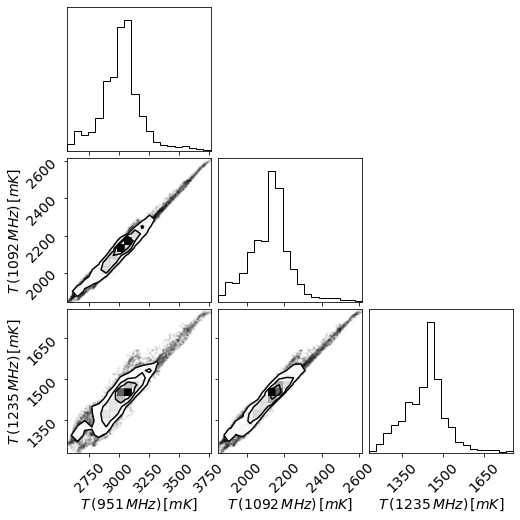} \\
    \caption{Corner plots to show the correlation across frequency for the `observed' data from the {\it{top:}} PSM simulations and {\it{bottom:}} GSM simulations.}
    \label{fig:corners}
\end{figure}

Our foreground simulations cover a 3600 square degree region centred at high Galactic latitude: $(l,b) = (120^{\circ}, -59^{\circ})$. We project the simulations onto the flat sky using the HEALPix Cartesian projection. Given the size of the survey area, using the flat-sky approximation does result in a level of angular distortion of the foreground maps, which would need to be corrected in an observational study. For our simulated study, however, we neglect this correction, instead choosing to consider the distorted projected maps as the `true' foreground model here. While the PSM simulations are just the summation of various components that each follow a power law, the GSM can be seen to display more complex frequency behaviour. Fig.~\ref{fig:gsmpsm} shows the mean emission temperatures from our region of interest at each frequency, multiplied by frequency squared and normalised to 1 at 956.7\,MHz, from the GSM and PSM foreground simulations. The PSM simulation can be seen to closely follow the form of a power law, while the GSM simulation displays a different spectral dependence. In Fig.~\ref{fig:corners} we show corner plots
for three frequencies of the total foregrounds within our region of interest plus 21cm signal data, convolved with the beam, for the cases of both PSM and GSM foregrounds. Both cases show a strong positive, linear correlation between frequency channels, but the GSM-based simulated data exhibit more distinct deviations from this linear correlation. Each GSM frequency map is accompanied with an error budget of 5 to 15\%, which would account for these deviations. This provides an interesting test case, as the GSM foreground model can be used as way to include analogous effects to spectral calibration errors alongside the standard perfect power-law representations of both the Galactic and extragalactic foregrounds in the PSM model. 

In this work, we use a 3600 square degree region centred at high Galactic latitude: $(l,b) = (120^{\circ}, -59^{\circ})$. As discussed in the previous section, the simulated data cubes, including foregrounds, are convolved with a realistic anisotropic beam model with sidelobes and non-trivial spectral structure. While we do not explicitly include a model for gain calibration errors, we use the GSM to provide the total foreground emission at each frequency and these maps are accompanied with a 5\% to 15\% error budget that is analogous to gain calibration errors. We do not consider additional spectral structure due to polarisation leakage in this paper, however.

\subsection{Mock galaxy distribution}
\label{sec:mockgalaxy}

In addition to the 21cm field, we also generate a mock galaxy catalogue from our simulated density field to allow us to study the impact of foreground cleaning on the galaxy-21cm cross-correlation signal too. To generate the catalogue, we use a log-normal transform of the linear matter density with a simple linear bias term $b(z)$ that depends only on redshift, so that the galaxy density contrast $\delta_g$ in each voxel is \citep{Agrawal:2017khv, nbody}
\be
1 + \delta_g(\vec{x}) = \frac{\exp \left ( b(z_c) \delta_m(\vec{x}) \right )}{\langle b(z_c) \delta_m(\vec{x}) \rangle},
\ee
where $z_c$ is the central redshift of the simulation box and in this case the angle brackets $\langle \dots \rangle$ denote a spatial average, necessary to obtain the correct normalisation. The galaxy counts are realised using a Poisson random draw in each voxel, with a rate parameter (expected number of galaxies in the voxel)
\be
\bar{N}(\vec{x}) = \delta V\, \bar{n}(z_c)\, (1 + b(z_c) \delta_m),
\ee
where $\delta V$ is the comoving voxel volume and $\bar{n}$ is the expected comoving number density of galaxies. This simplified treatment is sufficient to provide a reasonable model of galaxy-21cm cross-correlations for an auto-correlation survey; the linear bias model is expected to be accurate on large scales, and small-scale features of galaxy clustering that we have neglected, such as non-linear velocity dispersion, scale-dependent bias, and decorrelation between the 21cm and galaxy fields on small scales, are relatively less important due to the suppression of the 21cm field at high $k$ due to the instrumental beam. Note that there are exceptions to this reasoning for auto-correlation experiments with large dishes operating at low redshifts, such as the Parkes survey, which have sufficient angular resolution for small-scale galaxy clustering to be an important consideration \citep{Wolz:2017rlw, Anderson:2017ert}.

Rather than seeking to model a particular spectroscopic galaxy survey, we make simplistic choices of $b(z) = \sqrt{1+z}$ and $n(z) = 2 \times 10^{-3}\, {\rm Mpc}^{-3}$. This ensures that shot noise is unimportant, so the dominant effect on the cross-correlation signature will be systematic effects on the 21cm signal, such as the instrumental beam and residual foregrounds.

\section{Results}
\label{sec:results}

\begin{figure}
	\includegraphics[width=0.94\columnwidth]{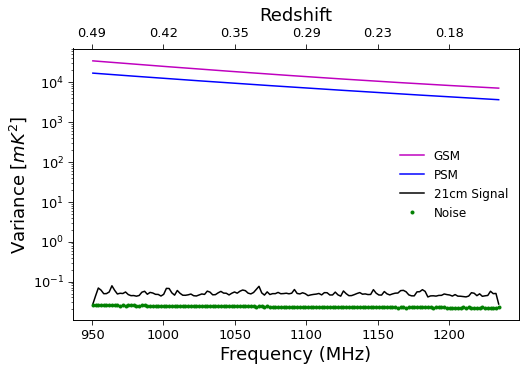} \\
	\includegraphics[width=0.94\columnwidth]{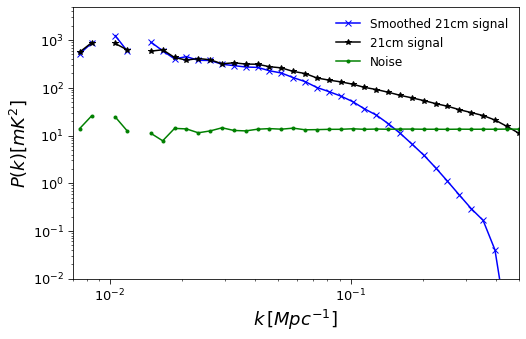} \\
    \caption{{\it{Top:}} The 21cm, PSM, and GSM foreground and noise variances as a function of frequency/redshift. {\it{Bottom:}} The radial power spectrum for the 21cm signal, before and after convolution with the MeerKAT beam, and instrumental noise.}
    \label{fig:var}
\end{figure}

\begin{figure}
	\includegraphics[width=0.94\columnwidth]{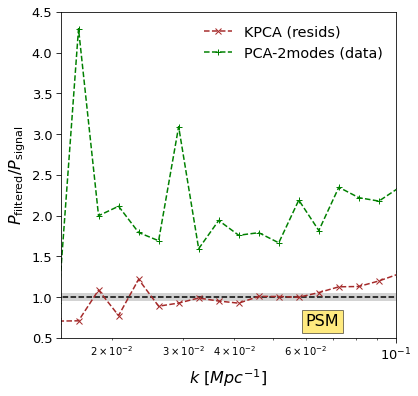} \\
	\includegraphics[width=0.92\columnwidth]{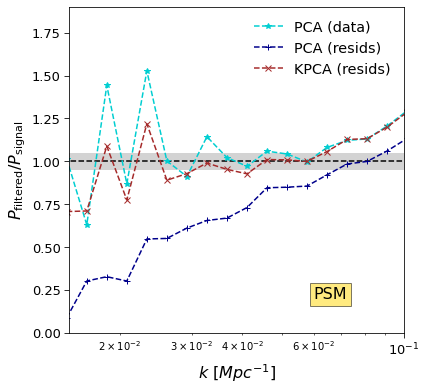}
	\includegraphics[width=0.92\columnwidth]{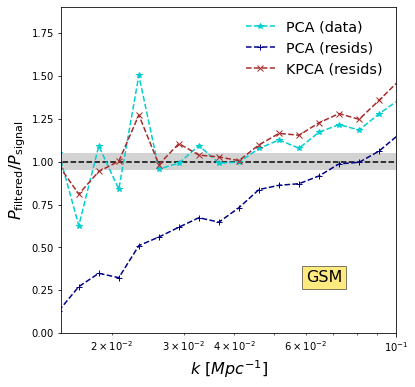}
    \caption{Ratio of recovered 1D power spectrum after foreground-filtering the data, $P_{\rm filtered}(k)$, to the input 21cm signal power spectrum, $P_{\rm signal}(k)$, for three different foreground filtering algorithms. The top two figures apply to the case of PSM foregrounds while the bottom figure is for the case of GSM foregrounds. The grey shaded area denotes the $\pm 5\%$ region around perfect signal recovery (a ratio of unity, denoted by the horizontal dashed line). The parentheses indicate whether the foreground cleaning method has been applied to the `observed' data cube (`data') or the pre-cleaned observation cube (`resids', i.e. following a 2-mode PCA removal).}
    \label{fig:res1}
\end{figure}

In this section, we present the results for the Kernel PCA foreground removal method applied to our total emission data cube. This data cube consists of the 21cm signal plus foreground components, which are convolved with the MeerKAT beam and then combined with Gaussian instrumental noise. The top panel of Fig.~\ref{fig:var} shows the variance of each of these components, before convolution with the MeerKAT beam, as a function of frequency. The effect of the beam on the 21cm radial power spectra can be seen in the bottom panel, where it causes the 21cm signal to dip below the noise level at large $k$ values. We restrict our foreground cleaning analysis to $k$ ranges lower than 0.1\,Mpc$^{-1}$ so as to ensure that the thermal noise level has negligible effects on our results. 

Kernel PCA is used in a number of different simulated scenarios, and for a range of different hyperparameters. In all cases we compare with PCA and ICA filters, using what we have found to be their optimal configurations in terms of recovery of the 1D signal power spectrum from the simulated data. For PCA and ICA, this means a choice of $n = 3$ foreground modes were selected for removal; this number produced the foreground-cleaned power spectra closest to the simulation 21cm power spectra convolved with the beam. For all of the analyses shown here, ICA behaves identically to PCA,\footnote{Very similar results from PCA/SVD and ICA have also been obtained in past simulated analyses \citep[e.g.][]{alonso}. When applied to real data, the results can be quite different however \citep[e.g.][]{Wolz:2015lwa}. This discrepancy may result from systematics present in the real data, or perhaps more strongly non-Gaussian foregrounds than are included in the simulations.} and so we only show the results from PCA in most figures. For all plots and tables where ICA is not explicitly mentioned, the term `PCA' should be taken to refer to `PCA and ICA'.

\begin{figure}
	\includegraphics[width=0.89\columnwidth]{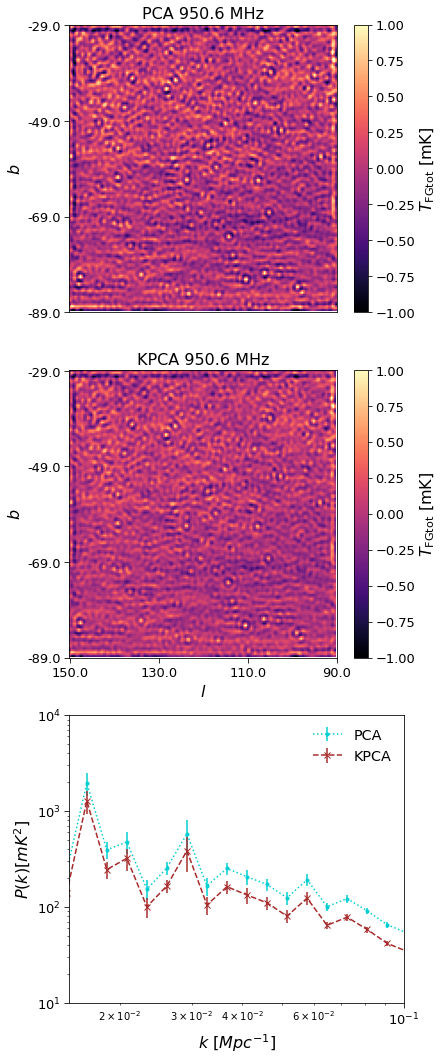}
    \caption{The foreground component after the initial pre-clean as identified by PCA ({\it{upper panel}}) and KPCA ({\it{middle panel}}) for the PSM foreground case. The maps are shown in Galactic coordinates. ({\it Lower panel}): The 1D power spectra for the KPCA and PCA foreground components, after the initial pre-clean.}
    \label{fig:pcacomps1}
\end{figure}

\begin{figure}
	\includegraphics[width=0.855\columnwidth]{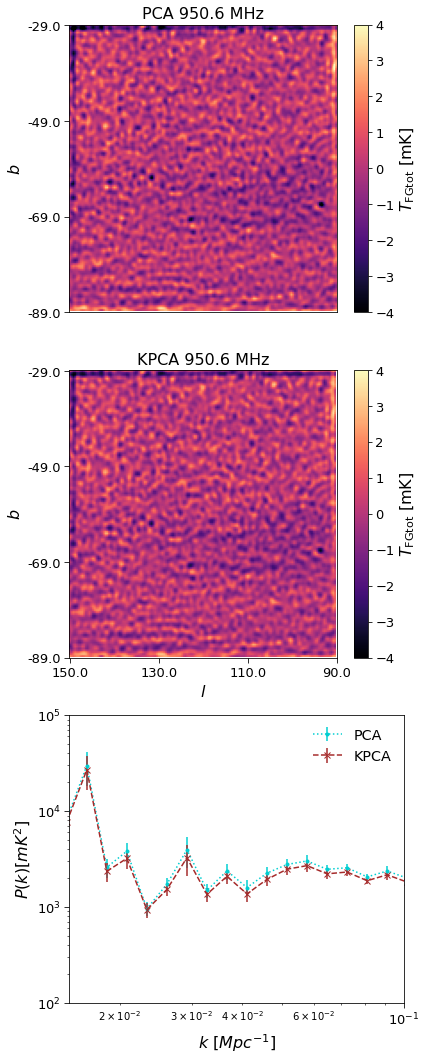}
    \caption{The same quantities as in Fig.~\ref{fig:pcacomps1}, but now for the GSM foregrounds.}
    \label{fig:pcacomps2}
\end{figure}

Kernel PCA requires three choices: 1) which kernel to use; 2) the number of modes, $n$, to be removed; and 3) the values of the kernel hyperparameters. Additionally, the method and degree of pre-cleaning must also be chosen. In this work we find the Sigmoid kernel to work best (see Section~\ref{sec:kernel}) and so must further choose $n$ and the values for the Sigmoid kernel's two hyperparamters: the kernel width $\gamma$ and the kernel offset $c$. As the kernel width is typically set to some scaling of $n$, we consider the true hyperparameter choice to be this scaling factor, $\epsilon$, where $\gamma = 1 / (\epsilon\, n)$.

For our standard set of simulations we choose a resolution of 150 by 150 pixels to capture the 60 by 60 degree field of view. In Section~\ref{sec:reso} we shall investigate the effect resolution choice has on our analysis. For our standard (150 by 150 pixels) simulations we set the number of modes to be removed by KPCA to be $n=5$ for the PSM foreground simulations and 7 for the GSM simulations. The fact that a larger number of modes was required to clean the GSM simulations reflects the additional spectral features present in the GSM foreground simulations. The optimal hyperparameter values will be discussed in more detail in Section~\ref{sec:reso}, but for now we shall just state the configuration used for our standard simulations: $\epsilon = 10$ and $c = 0.9$. 

For our analysis we compare the recovered 21cm emission from each cleaning method with the true 21cm signal convolved with the frequency-dependant beam. The ratio between the recovered PCA/ICA/KPCA-filtered 1D power spectrum and the true 21cm beam-convolved 1D power spectrum is used as the figure of merit throughout. As the ideal cleaning method would produce a 1d power spectrum ratio of 1, the ratio = $1 \pm 0.05$ region is highlighted on each of the ratio plots. 

Fig.~\ref{fig:res1} shows the ratio of the recovered to true 1D power spectrum for each investigated technique. The top panel demonstrates the role of pre-cleaning for the PSM foregrounds, with the power spectrum ratio for the residual produced by PCA pre-cleaning shown in green. This residual is roughly double the level of the 21cm signal. The brown curve shows the result of applying KPCA to this residual, as opposed to the total data cube. The middle panel shows the same brown curve, but this time plotted alongside the result from a PCA (3-mode) clean on the total data (cyan) and the residual from before (blue). Using PCA first with 2 modes, then with 3, is no different from using PCA with 5 modes. Unlike Kernel PCA, ICA and linear PCA end up over-cleaning the total data and removing a substantial fraction of the 21cm signal when given the pre-cleaned data cube. KPCA, on the other hand, makes use of the this pre-cleaning step, and can be seen to outperform both ICA and PCA after the data have been pre-cleaned. The bottom panel of Fig.~\ref{fig:res1} also shows the ratio of the recovered to true 1D power spectrum but for the case of GSM foregrounds, with similar conclusions.

The advantage of using KPCA over PCA is far more pronounced at low $k$/larger scales. In fact, it can seen for the GSM foreground simulations that PCA outperforms KPCA at high $k$ values. Comparing both the middle and bottom plot of Fig.~\ref{fig:res1} reveals that gain calibration-like errors, as modelled by the additional spectral structure in the GSM simulations, result in larger inaccuracies for PCA, ICA, and KPCA at small scales (high $k$ values). For $k$ values lower than $6\times 10^{-2}\,{\rm Mpc}^{-1}$, KPCA is seen to outperform PCA, but for the smallest scales PCA would remain the preferred method. The inability to completely recover the 21cm signal at large angular scales/low $k$ values has been shown to be a feature of blind cleaning techniques \citep{pcalim}; the number of spurious correlations between foregrounds and the 21cm signal is inversely proportional to the square root of the number of independent signal modes (which decreases at lower $k$ values). Any improvements that can be made over standard PCA at large angular scales are therefore worth pursuing.

For both KPCA and PCA, the eigenvectors with the largest eigenvalues correspond to the components that the algorithm has, effectively, identified as foregrounds. We use KPCA on the pre-cleaned observation cube, which has the first two largest PCA components removed from it already, and so it is constructive to compare what both the KPCA and PCA determine to be a `foreground' {\it after} this initial pre-cleaning step. Fig.~\ref{fig:pcacomps1} and Fig.~\ref{fig:pcacomps2} compare these two post-pre-cleaning estimates of the `foreground' component for both the PSM and GSM foreground cases, respectively. The plots show the foreground estimate for the first frequency channel, while the power spectra take into account the full data cube and so all available frequencies.

Fig.~\ref{fig:pcacomps1} displays point-like emission contributions whilst Fig.~\ref{fig:pcacomps2} does not; this is because the GSM is a diffuse sky model, which does not directly include a resolvable point source contribution. There is little visible difference between the structure seen by the KPCA and PCA foreground estimate, after the pre-clean, in both the PSM and GSM cases. The most notable difference between the two in both foreground cases (though the difference is more pronounced for the PSM case) is the power level, which is always higher for PCA. KPCA, with the number of components set to 5, removes less power from the total emission maps than PCA with the number of components set to 3. A clear advantage to using KPCA over standard PCA is therefore the finer level of precision with which contaminants can be identified and removed. In some sense, this can be thought of as allowing fractional numbers of PCA components to be removed, e.g. subtracting the equivalent of more than 2 but less than 3 PCA components from the total observational cube. With appropriate optimisation of the KPCA parameters, this should allow for reduced signal loss while still removing the bulk of the foreground emission.

\begin{figure}
	\includegraphics[width=0.9\columnwidth]{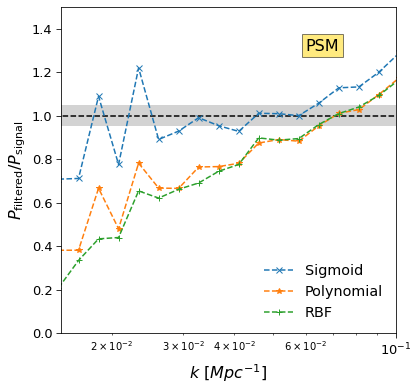} \\
	\includegraphics[width=0.9\columnwidth]{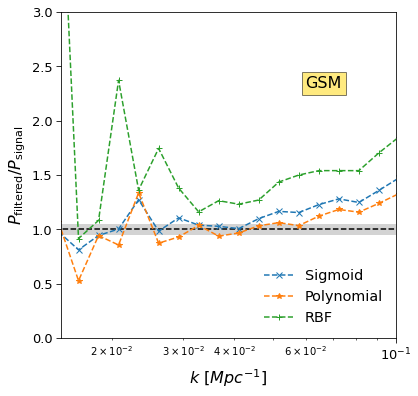} \\
    \caption{Ratio of recovered 1D power spectrum after foreground-filtering the data, $P_{\rm filtered}(k)$, to the input 21cm signal power spectrum, $P_{\rm signal}(k)$, for Kernel PCA using different kernels. Results are shown for the PSM ({\it{upper panel}}) and GSM ({\it{lower panel}}).}
    \label{fig:kernelchoice}
\end{figure}

\begin{figure*}
	\includegraphics[width=0.95\columnwidth]{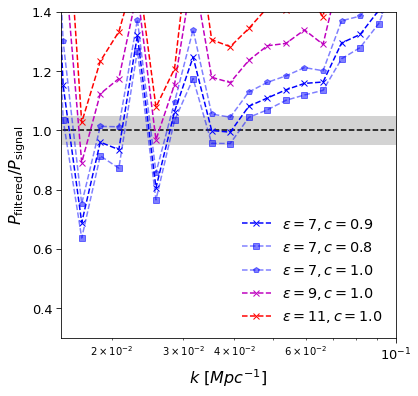}
	\includegraphics[width=0.95\columnwidth]{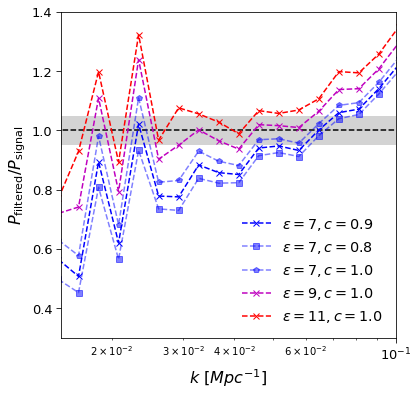}
    \caption{Ratio of recovered 1D power spectrum after foreground-filtering the data, $P_{\rm filtered}(k)$, to the input 21cm signal power spectrum, $P_{\rm signal}(k)$, for Kernel PCA using different values of $\epsilon$ and $c$. The resolution 100 results are shown in the left panel while the resolution 150 results are shown on the right. Both plots are for the case of PSM foregrounds.}
    \label{fig:res2}
\end{figure*}

\subsection{Effect of different choices of kernel}
\label{sec:kernel}

The choice of which kernel/hyperparameters to use has the greatest effect on the successful application of KPCA. Similar discussions can be seen in the application of Gaussian Process Regression to 21cm intensity mapping simulations \citep{mertens, Ghosh:2020fpy, kern, paula}; siting GPR within a Bayesian framework, the optimum choice of kernel and hyperparameters can be found by comparing the log-marginal likelihood (LML) for each kernel/hyperparameter set under investigation. In other applications of KPCA, a cross-validation technique is often applied where the various kernel/hyperparameter options are tested on subsets of the data, and the kernel/hyperparameters that can produce the most consistent reproductions of the desired image (or otherwise) from these subsets are adopted. In order to compare the results from different kernel selections without knowing the true form of the image that KPCA is trying to recover, the kernel estimates must be mapped back from their feature space into the common input space (the image pixel space). This is known as the `pre-imaging' problem \citep{pick}.

In our application of KPCA, we determine the optimal kernel/hyperparameter set empirically, which is also what we do for PCA and ICA. Future applications of KPCA to observed 21cm intensity mapping data could instead use simulations to inform the kernel/hyperparameter choice, as well as the technique of eigenvalue convergence which looks at the ratio between foreground eigenvalues and the total data eigenvalues \citep{steve}. We do not investigate optimisation methods further here however.

Our analysis mostly focuses on the Sigmoid kernel, the choice of which we justify in Fig.~\ref{fig:kernelchoice}. For both the PSM and GSM foreground cases, the sigmoid kernel is optimal for our particular simulation set-up when compared with the polynomial and RBF kernels. Interestingly, for the PSM foregrounds the sigmoid kernel produces a different recovered 21cm power spectrum shape as a function of $k$ when compared to the polynomial and RBF kernels, while for the GSM foreground all three kernels can be seen to seen to have similar recovered 21cm power spectrum forms. 

For the sake of comparison, each kernel is given the same set of shared hyperparameters ($n=5, \epsilon = 10, c = 0.9$). The polynomial kernel also has an additional parameter: the order of the polynomial, $d$. For the case of the PSM foreground, the optimal value for this parameter was $d = 1$, whilst the GSM foregrounds the optimal value was found to be $d = 4$. We examine the effect of changing hyperparameter values in the next section.

\subsection{Effect of the data cube resolution}
\label{sec:reso}
The results from the previous section are for a 60 degree by 60 degree region of the sky projected onto a Cartesian grid of dimensions 150 by 150 pixels, which provides a pixel size of $0.4^\circ$. \citet{wang21} use a pixel size of $\sim 0.3^\circ$ for the MeerKLASS survey, as this is a third of the average beam FWHM. In this section, we investigate the effect of the pixel resolution on the foreground cleaning techniques. The simulations continue to be convolved with the MeerKAT L-band beam, but for these results the Cartesian grid is changed from 150 pixels squared to 100 ($0.6$ degree pixel size) and then 200 ($0.3$ degree pixel size). Note that the simulation resolution is specified in three dimensions, and so by changing the number of angular pixels we also change the number of frequency channels e.g. from 150 to 100 or 200. As such, we are really testing the dependence of the KPCA method on the spatial {\it and} spectral resolution here. It should also be noted that changing the frequency bandwidth also alters the level of thermal noise, increasing it for lower resolutions and decreasing it for higher resolutions. Even at our lowest resolution of 100 frequency channels, the thermal noise power level is still below the smoothed 21cm power level for $k < 0.1$~Mpc$^{-1}$ however.

\renewcommand{\arraystretch}{1.4}
\begin{table*}
 \begin{tabular}{lllccclc}
 \cline{1-8}
 {\bf FG model} & {\bf{Resolution}} & {\bf{$k$ range (Mpc$^{-1}$)}} & {\bf PCA ($n=3$)} & {\bf ICA ($n=3$)} & {\bf KPCA} & {\bf Sigmoid kernel params.} & {\bf Benefit} \\
 \cline{1-8}
\multirow{6}{*}{PSM} & \multirow{2}{*}{100} & $[0.015, 0.058]$ &  0.14 & 0.14 & 0.14 & \multirow{2}{*}{$n=5, \epsilon = 5, c=0.9$} & = \\
& & $[0.058, 0.1]$ & 0.20 & 0.20 & 0.18 &  & \cmark \\
\cline{2-8}
    & \multirow{2}{*}{150} & $[0.015, 0.058]$ &  0.17 & 0.17 & 0.10  & \multirow{2}{*}{$n=5, \epsilon = 10, c=0.9$} & \cmark  \\
    & & $[0.058, 0.1]$ & 0.11 & 0.11 & 0.10 &  & \cmark \\
\cline{2-8}
    & \multirow{2}{*}{200} & $[0.015, 0.058]$ & 0.07 & 0.07 & 0.05  & \multirow{2}{*}{$n=5, \epsilon = 20, c=0.9$} & \cmark  \\
    & & $[0.058, 0.1]$ & 0.07 & 0.07 & 0.09 & & \xmark \\
\cline{1-8}
\multirow{6}{*}{GSM} & \multirow{2}{*}{100} & $[0.015, 0.058]$ &  0.14 & 0.14 & 0.44 & \multirow{2}{*}{$n=7, \epsilon = 5, c=0.9$} & \xmark \\
& & $[0.058, 0.1]$ &  0.30 & 0.30 & 0.87 & & \xmark \\
\cline{2-8}
    & \multirow{2}{*}{150} & $[0.015, 0.058]$ &  0.14  & 0.14  & 0.09  & \multirow{2}{*}{$n=7, \epsilon = 10, c=0.9$} & \cmark \\
    & & $[0.058, 0.1]$ & 0.19 & 0.19  & 0.25 & & \xmark \\
\cline{2-8}
    & \multirow{2}{*}{200} & $[0.015, 0.058]$ &  0.08 & 0.08 & 0.05 & \multirow{2}{*}{$n=7, \epsilon = 20, c=0.9$} & \cmark \\
    & & $[0.058, 0.1]$ & 0.11 & 0.11 & 0.17 & & \xmark \\
 \cline{1-8}
 \end{tabular}
 \caption{The mean absolute deviation from unity of $P_{\rm filtered} / P_{\rm signal}$ for each foreground cleaning method used in the cases of GSM and PSM foreground simulations. The average is taken over the $k$-bins in the range specified. The `Benefit' column highlights whether or not any advantage over PCA was seen from the use of KPCA. Note that a 2-mode PCA pre-cleaning step has been applied to the data only for the KPCA case.} 
 \label{table:resmads}
\end{table*}

Fig.~\ref{fig:res2} explores the effect of the $\epsilon$ and $c$ hyperparameters on the power spectrum results for both the 100 and 150 pixel resolution simulations using the PSM foregrounds. $\epsilon$ values of 7, 9, and 11 (with $c=0.9$) are shown alongside results where we fix $\epsilon = 7$ and vary $c$ with values 0.8, 0.9, and 1.0. For the resolution 150 case, the choice of $\epsilon = 10$ is optimal, but for the lower resolution simulation, decreasing $\epsilon$ (increasing $\gamma$) improves the performance of Kernel PCA. We find the optimum $\epsilon$ values to be 5, 10, and 20 for the 100, 150 and 200 pixel resolutions respectively. This is somewhat understandable; as $\epsilon$ is increased, $\gamma$ is decreased, leading to a broader decision region in feature space \citep{gamma}. As more pixels are present to describe the same spatial and spectral information in the higher resolution cases, one can imagine a broader hypersurface being drawn around these data points in feature space. The same response to increasing $\epsilon$ is seen for the GSM foregrounds as for the PSM foregrounds, so we do not present the additional plots for the GSM case.

\begin{figure}
	\includegraphics[width=0.98\columnwidth]{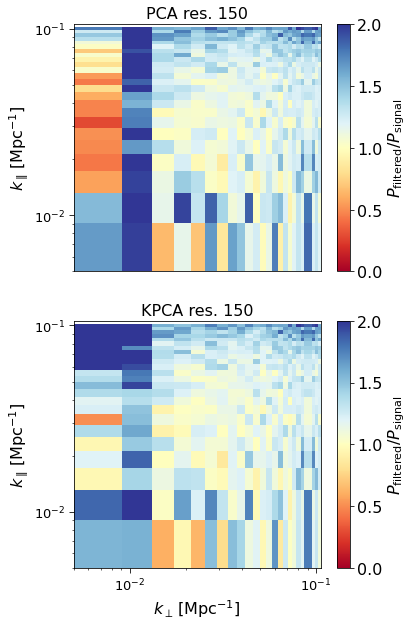}
    \caption{Ratio of recovered 2D power spectrum after foreground-filtering the data using KPCA or PCA, $P_{\rm filtered}(k_\perp, k_\parallel)$, to the input 21cm power spectrum, $P_{\rm signal}(k_\perp, k_\parallel)$, for the resolution 150 case (GSM foregrounds). Blue regions denote an excess of power, implying the presence of residual foregrounds, while red regions imply over-subtraction has occurred.}
    \label{fig:pk2d150}
\end{figure}

\begin{figure}
	\includegraphics[width=0.98\columnwidth]{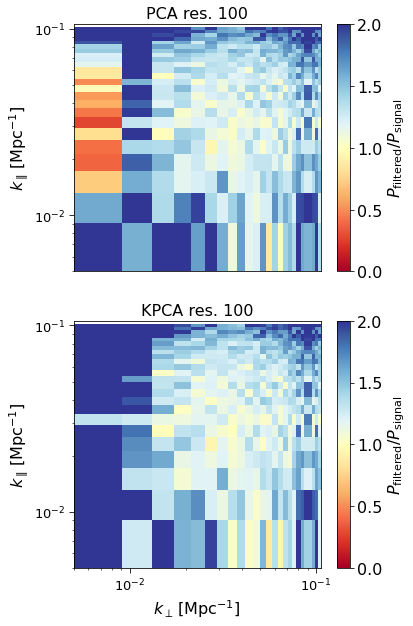}
    \caption{The same quantities as in Fig.~\ref{fig:pk2d150}, but now for resolution 100.}
    \label{fig:pk2d100}
\end{figure}

The combination of the effects of $\epsilon$ and $c$ is particularly interesting, as it can be seen that through careful manipulation of both these parameters, an increasingly faithful representation of the 21cm signal 1D power spectrum can be obtained. This is both an advantage and disadvantage of KPCA -- the hyperparameters allow for a more sensitive tuning of the component separation than would be allowed by linear PCA for example, which enables KPCA to perform between 5 and 10\% better than linear PCA (except at the smallest scales). Non-optimal hyperparameters lead to an equivalent performance between KPCA and linear PCA however, and so the ultimate effectiveness of the method depends on the tuning that has been applied.

Table~\ref{table:resmads} presents the mean absolute deviation (about the fiducial value of unity) of $P_{\rm filtered} / P_{\rm signal}$ for each foreground cleaning method, where the average is taken over $k$. The lower the deviation, the more reliably effective the method is at recovering the 21cm signal on average. The deviation is calculated for two ranges of $k$ to highlight the difference in behaviour as a function of scale. The hyperparameters used for each scenario are stated alongside the results.

For the PSM simulation, KPCA is seen to offer an improvement over PCA and ICA at all resolutions and angular scales except for the smallest scales at the highest resolution tested (200 pixels). For the GSM simulation however, KPCA is only seen to outperform linear PCA and ICA at large scales for the 150 and 200 pixel resolution simulations, and not across any scales for the 100 pixel resolution simulation.

Fig.~\ref{fig:pk2d150} shows the cylindrically-averaged 2D power spectrum ratio between the recovered and true signal for PCA (top panel) and KPCA (bottom panel) in the case of the GSM foregrounds at 150 pixel resolution. In the 2D power spectrum case, $k_\perp$ is equal to $\sqrt{k_{x} + k_{y}}$ and represents angular scales, while $k_\parallel$ corresponds to the frequency/redshift direction. The 3D band powers have been binned into 25 equally spaced bins in $k_\perp$ between 0.005 and 0.11 Mpc$^{-1}$, and the contributing $k_\parallel$ powers have been limited to those between 0.005 and 0.11 Mpc$^{-1}$. Both methods can be seen to struggle to recover the 21cm signal at the largest angular scales (smallest $k_\perp$ values) across all $k_\parallel$ values, with PCA showing the largest signal loss. However at 100 pixel resolution (Fig.~\ref{fig:pk2d100}), KPCA instead overestimates the 21cm signal at the largest angular scales across all $k_\parallel$ values.

KPCA (with a Sigmoid kernel) does not appear to be well-equipped to cope with the additional spectral structure present in the GSM simulation when the data are under-sampled (a pixel size of 0.6 degrees for an average beam FWHM of 1 degree is only one pixel per beam), or for any resolution at small scales (high $k$ values). Note however that the polynomial kernel case shown in the bottom panel of Fig.~\ref{fig:kernelchoice} does appear to perform slightly better on small scales for the GSM-based simulations, and so better performance on small scales can likely be attained with a different choice of kernel and hyperparameters.

\subsection{Performance in the presence of a mask}

\begin{figure}
	\includegraphics[width=1.\columnwidth]{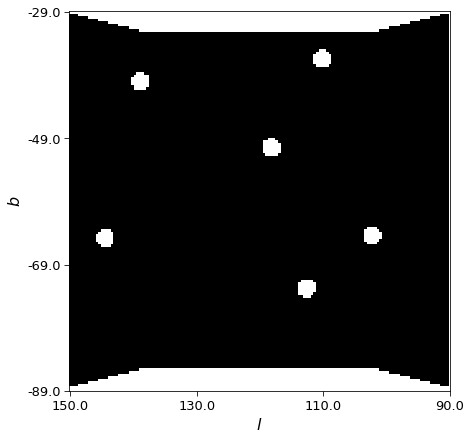} \\
    \caption{Point source and scan strategy mask applied at each frequency, shown in Galactic coordinates. The mask is intended to model the excision of a few bright point sources in the field (with masked regions of order the beam width), and irregular survey edges due to the choice of scan strategy.}
    \label{fig:mask}
\end{figure}

Incomplete data coverage, e.g. due to point source masking and scanning strategies that lead to irregular survey areas, is an important consideration when performing foreground removal. In this section we introduce a mask, shown in Fig.~\ref{fig:mask}, to mimic the effects of masking bright point sources and a non-square scanning strategy. We construct the mask under the assumption that the Cartesian projection at each frequency is in fact the measured temperature map, i.e. ignoring the distortions to the mask that would be introduced by performing a projection from spherical coordinates. The mask is applied to each observed frequency, after which we reapply our foreground cleaning techniques. To soften the sharp features introduced in the data by masking, we in-paint the masked regions using the average values of their nearest-neighbour pixels.

\begin{figure}
	\includegraphics[width=1.\columnwidth]{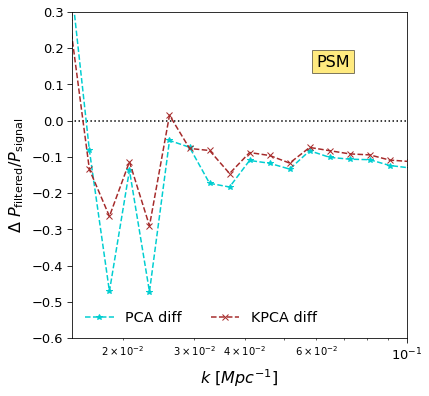} \\
		\includegraphics[width=1.\columnwidth]{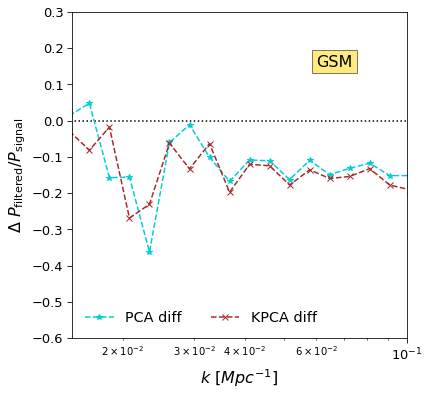} \\
    \caption{Change in the ratio of the recovered-to-true 1D power spectrum between foreground-filtering the masked data and the full data. Results are shown for the PSM ({\it{upper panel}}) and GSM ({\it{lower panel}}). A black dotted zero-line has been added to guide the eye.}
    \label{fig:resmask}
\end{figure}

In Fig.~\ref{fig:resmask} we plot the change in the ratio of the recovered-to-true 1D power spectra from filtering the full observational data to filtering the masked data set. At the smallest $k$ values, for the PSM foregrounds only, PCA can be seen to perform substantially worse in the presence of the mask, more so than KPCA. Masking and in-painting the data degrades the accuracy of the 21cm estimates produced by all the techniques tested, but no single method is worse affected by the inclusion of a mask across all scales. The same conclusion can be drawn from both the PSM and GSM foreground simulations. Arguably a more sophisticated method of in-painting than just a nearest-neighbour average could have been used, but the only purpose of this test was to ensure that masking would not degrade the performance of KPCA more so than it would PCA/ICA.

\subsection{Smoothing to a common resolution}
A technique sometimes used to mitigate the effects of the frequency-dependant beam, its sidelobe structure, and polarisation leakage, is to smooth the data at all frequencies to a common angular resolution. For example, \citet{tfunc2} used a common resolution of 1.4 times the largest beam size in their analysis. Since the beam is typically largest at the lowest frequencies, this degrades the angular resolution of the data at higher frequencies, losing some information from the 21cm field in the transverse direction at lower redshifts. In some applications, this loss is acceptable given the additional mitigation of systematic effects, whereas in others \citep[e.g. BAO recovery from the correlation function;][]{Kennedy:2021srz, Avila:2021wih}, any amount of additional angular smoothing is undesirable.

To investigate the effect of smoothing on the performance of KPCA, we smoothed the simulated data to a common angular resolution of $2.25^\circ$, which corresponds to $1.4\times$ the maximum beam size (the beam FWHM ranges from approximately $1.24^\circ - 1.61^\circ$ in our simulations). Practically, this entails taking the combined 21cm and foreground datacubes that have already been convolved with the frequency-dependant {\tt katbeam} beams, plus noise, and then convolving the map in each frequency channel with a Gaussian beam of FWHM $= \sqrt{(2.25^\circ)^{2} - \theta_{{\rm{approx}}}^{2}}$, where $\theta_{{\rm{approx}}} \approx 1.2 \lambda / (13.5 {\rm m})$ is the FWHM at each frequency given by the Gaussian approximation for the MeerKAT dishes. This procedure accounts for the existing degree of smoothing at each frequency, and then applies whatever additional smoothing is necessary to reach the common angular resolution. The Gaussian approximation for the existing beam smoothing does not take into account features such as sidelobes however, and so the resulting `effective' beam after smoothing will not necessarily be completely frequency-independent.

An $n=2$ PCA pre-cleaning step was used for KPCA, as before, and the reference PCA and ICA methods continued to remove $n=3$ modes. The hyperparameters of KPCA, however, needed to be readjusted to perform optimally in this new setup. For the original resolution PSM and GSM simulation data, the best-performing hyperparameters of the Sigmoid kernel were $n=5, \epsilon=10, c=0.9$ and $n=7, \epsilon=10, c=0.9$ respectively. For the smoothed data, the hyperparameters were set to $n=3, \epsilon=5, c=0.9$ for the PSM foreground simulations. The reduction in the number of modes removed, $n$, and the kernel width scaling factor, $\epsilon$, reflects the simpler spectral structure of the data due to the smoothing. For the GSM foregrounds, however, the Sigmoid kernel was no longer the optimal choice, and so we switched to a polynomial kernel with hyperparameters $n=7, \epsilon=10, c=0.9, d=4$, which provided a better 21cm power spectrum recovery.

\begin{figure}
	\includegraphics[width=1.0\columnwidth]{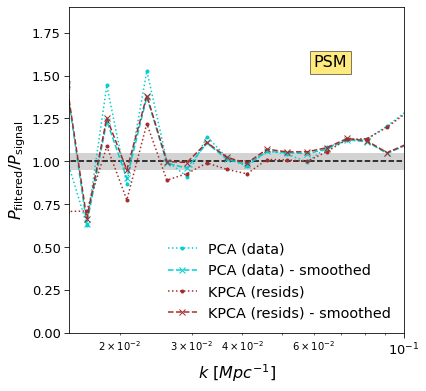} \\
	\includegraphics[width=1.0\columnwidth]{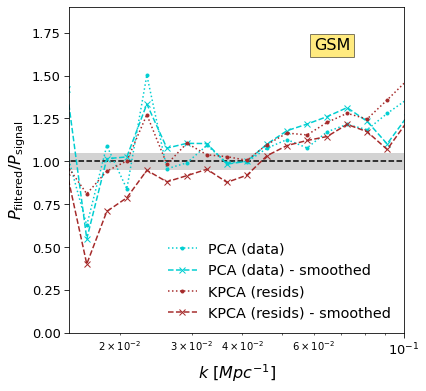} \\
    \caption{Ratio of recovered 1D power spectrum after foreground-filtering the data, $P_{\rm filtered}(k)$, to the input 21cm signal power spectrum, $P_{\rm signal}(k)$ for the data at their original resolution and also in the case where the data have been smoothed to a common resolution. Results are shown for the PSM ({\it{upper panel}}) and GSM ({\it{lower panel}}) simulations.}
    \label{fig:comres}
\end{figure}

Fig.~\ref{fig:comres} shows the 1D power spectrum recovered-to-true ratio for both the PSM and GSM foreground simulations where the `observed' data have been smoothed to a common resolution of $2.25^\circ$. The power spectrum ratios for the original resolution data, previously shown in Fig.~\ref{fig:res1}, have also been re-plotted for comparison. For the PSM simulations it can be seen that the performance of PCA/ICA improves slightly after smoothing, particularly on large scales, whereas the performance of KPCA degrades slightly, also mostly on large scales. In fact, the PCA/ICA and KPCA methods behave almost identically after smoothing, suggesting that whatever information KPCA was taking advantage of has been removed by the smoothing step. The differences are relatively minor however, in-keeping with results from previous sections that show that KPCA typically offers only moderate improvements over PCA.

The picture is slightly more nuanced in the case of the GSM foregrounds. PCA/ICA do slightly worse after smoothing, particularly on smaller scales, although there are some differences in behaviour on large scales too. Conversely, KPCA (now with a polynomial kernel) performs slightly better on small scales after smoothing, but worse on large scales. The PCA/ICA and KPCA results are not now almost identical after smoothing, as they were for the PSM foreground model, which may be due to either the additional spectral structure in the GSM model or the different choice of KPCA kernel.

In either case, it is clear that smoothing has an impact on the ability of KPCA to recover the 21cm signal, tending to reduce its advantage compared with PCA/ICA, albeit in a scale-, kernel-, and foreground model-dependent manner. In contrast, smoothing can help improve the performance of PCA/ICA. This suggests that applications of these methods to real data should use unsmoothed data for KPCA and smoothed data for PCA/ICA. (Also recall our discussion of resolution effects in Sect.~\ref{sec:reso}.)

\subsection{Cross-correlations with galaxy surveys}

\begin{figure}
	\includegraphics[width=1.\columnwidth]{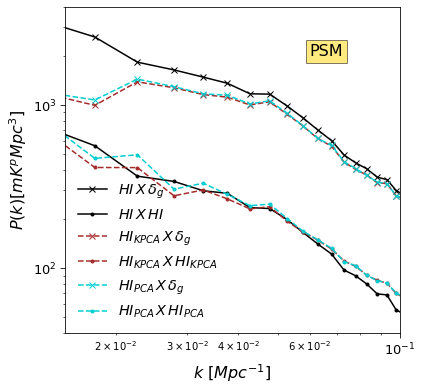} \\
	\includegraphics[width=1.\columnwidth]{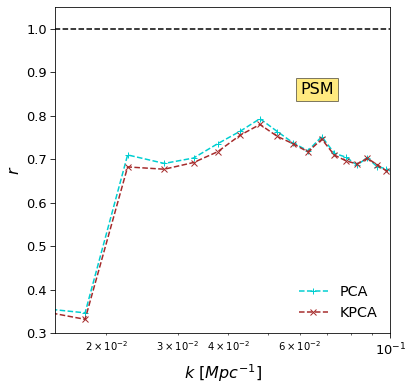}
    \caption{1D cross- and auto-correlation power spectra ({\it{upper panel}}) and cross-correlation coefficient ({\it{lower panel}}) between the KPCA/PCA foreground-cleaned maps and the galaxy sample data  for the PSM simulation. The y-axis units in the upper panel (${\rm mK}^{p}$) are $p=1$ for the cross-correlation spectra and $p=2$ for the auto-correlation spectra.}
    \label{fig:coors1}
\end{figure}

Whilst the detection of the 21cm signal from auto-power spectra relies heavily on the quality of the foreground cleaning, the use of cross-correlations with galaxy catalogues is a way to break this dependence \citep{steve19, pad}. Optical galaxy surveys resolve individual galaxies, providing a high-resolution counterpart to intensity maps in overlapping volumes. To determine the impact of foreground cleaning on the 21cm-galaxy cross-correlation power spectra, we also calculate the cross-power between the foreground-cleaned data cubes and the mock galaxy catalogue in the same simulated volume described in Sect.~\ref{sec:mockgalaxy}. We use the same kernel/hyperparameter combinations as in Fig.~\ref{fig:res1}, with the PCA method applied directly to the data and the KPCA method applied to the pre-cleaning residuals.

\begin{figure}
	\includegraphics[width=1.\columnwidth]{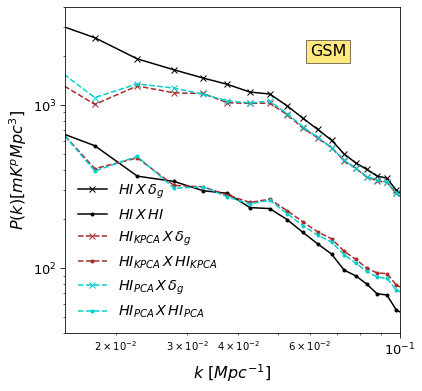} \\
	\includegraphics[width=1.\columnwidth]{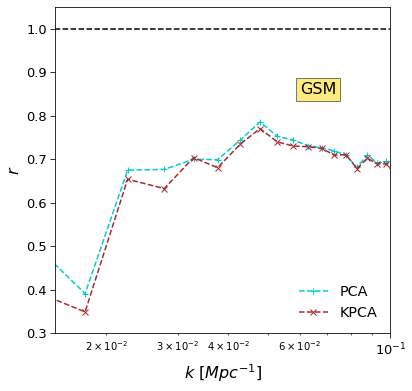}
    \caption{The same quantities as in Fig.~\ref{fig:coors1}, but now for the GSM foregrounds.}
    \label{fig:coors2}
\end{figure}

The cross-correlation power spectra are computed between fluctuations in the 21cm brightness temperature and the galaxy number counts. The top panels of Fig.~\ref{fig:coors1} and Fig.~\ref{fig:coors2} show the 1D cross- and auto-correlation power spectra for our foreground cleaned maps for the cases of the PSM and GSM foregrounds. The 21cm signal (convolved with the MeerKAT beam) auto-correlation spectrum is shown for reference, as is the 21cm-galaxy cross-correlation spectrum and the auto-correlation power spectrum results first shown in Fig.~\ref{fig:res1}. The bottom panels show the cross-correlation coefficient between the foreground cleaned maps and the galaxy sample data, defined as:
\begin{equation}
r = \frac{P_{\times}(k)}{\sqrt{P_{{\rm{21cm}}} (k) \, P_{{\rm{halo}}} (k)}},    
\end{equation}
where $P_{\times}(k)$ is the cross-correlation power spectrum for the foreground cleaned map and the galaxy sample data, $P_{{\rm{21cm}}} (k)$ is the auto-correlation power spectrum for the 21cm signal and $P_{{\rm{halo}}} (k)$ is the auto-correlation power spectrum for the galaxy sample data. For the case of PSM foregrounds, the cross-correlation power spectra from the KPCA and PCA recovered 21cm signal are quite similar; there is no clear advantage from choosing one method over the other, except for a slight reduction in signal loss from PCA on the largest scales. For the case of GSM foregrounds, PCA is also seen to sustain slightly less signal loss than KPCA at the smallest $k$ values in cross-correlation. It is likely that the signal loss in the cross-power for the KPCA method could be reduced, e.g. by choosing hyperparameters that increase the residual foregrounds in the auto-power.

\section{Conclusions}
\label{sec:conclusions}

Foreground emission is the major source of systematic errors in 21cm intensity maps, both as a contaminant in its own right, and through difficult-to-model interactions with the spectrally-complex instrumental response. If foregrounds can be cleaned from the data efficiently enough, 21cm auto-power spectrum analyses can be conducted in a manner analogous to spectroscopic galaxy surveys \citep[e.g.][]{Soares:2020zaq, Kennedy:2021srz, Avila:2021wih} for example, enabling a wide variety of cosmological applications for this tracer \citep[e.g.][]{Chang:2007xk, Bull:2014rha}. Foregrounds are also important in 21cm-galaxy cross-correlation analyses \citep[e.g.][]{Villaescusa-Navarro:2014rra, steve19}, as while in principle they are (mostly) uncorrelated with galaxies, the additional variance in the data contributed by the foregrounds significantly increases the noise on the cross-power spectrum if they are not at least partially removed.

A wide variety of methods exist to remove foreground emission, each using different ways to separate the 21cm signal and the foregrounds, such as according to: signal-to-noise ratio, spectral smoothness, statistical independence, angular structure, sparsity, localisation in other bases and so on. With all currently known methods, there are trade-offs to be made between the efficiency of foreground removal and the risk of over-subtracting, such that some of the cosmological signal is lost too. The range of (Fourier) scales over which the signal can be reliably recovered is the key metric for cosmological analyses, but the sensitivity of the method to prior assumptions about the structure of the data, instrumental effects etc. must also be taken into account. Methods that work extremely well on simulated data may struggle with inevitably more complex real-world data, particularly if those methods have been extensively optimised under particular simulated model assumptions.

It therefore behooves us to apply a range of different foreground removal methods to real-world datasets, in the hope that consistent answers are obtained from methodologically very different approaches, thus improving our confidence in the analysis. A lack of convergence can hint at the presence of unmitigated systematic effects instead \citep{tfunc2, Wolz:2015lwa}.

In this spirit, we have investigated a method that is new to this domain -- Kernel PCA (KPCA). Unlike some other methods, KPCA seeks to represent the data in a {\it higher}-dimensional space than the data itself, using various non-linear combinations of the original dimensions of the data. This ultimately allows the data to be split into components with more complicated functional forms than allowed by purely linear decomposition methods like PCA, when said higher-dimensional functions are projected back down onto the original dimensions of the data. Thanks to the `kernel trick', the computational expense of constructing the KPCA model is limited, as the full higher-dimensional representation of the data is never needed. The allowed functional forms and their weightings are determined by the choice of kernel and its hyperparameters, which can be selected to promote structure such as localisation, radial symmetry and so forth. While KPCA is in principle a blind method, as with other methods (e.g. Gaussian Process Regression), the need to choose the kernel/hyperparameters means that some form of tuning can be applied that may benefit from knowledge of the data.

We have applied KPCA to simulations that incorporate a log-normal 21cm signal model, noise, instrumental effects (e.g. beams with sidelobes), and a choice of two foreground models with substantially different model assumptions. We have studied the performance of KPCA in several scenarios, with a variety of choices of kernel and hyperparameters. In all cases we have compared its efficacy to the more established (linear) PCA and ICA methods, which we have used as a benchmark. Our findings about KPCA as a foreground removal method are as follows:
\begin{itemize}
 \item Kernel PCA typically performs better than PCA or ICA in terms of auto-power spectrum recovery on intermediate Fourier scales when an initial pre-cleaning step is performed (e.g. first using PCA with a small number of modes to reduce signal loss).
 
 \item While it reliably improves on PCA for the Planck Sky Model foregrounds, KPCA behaves more erratically when applied to the GSM foregrounds, likely due to difficulties in handling the additional spatial/spectral structure that is present. KPCA also does worse than PCA with under-sampled (lower-resolution) data.
 
 \item KPCA does not display the same 21cm signal loss as PCA at low values of $k_\perp$, though for the GSM foregrounds it leaves significant foreground contamination on these large angular scales.  

 \item The choice of kernel has a substantial impact on the effectiveness of the KPCA method. We found the Sigmoid kernel to have the best properties overall, e.g. reducing the degree of signal loss on large scales. The method also depends on the choice of hyperparameters, but in a relatively smooth and controlled way.

 \item KPCA is robust to missing data and masking; the presence of a mask slightly degrades every method we studied, but in broadly the same way for each of them.
 
 \item Smoothing each frequency channel to a common beam resolution (e.g. to average out small-scale beam effects) is generally helpful for PCA, but slightly degrades the performance of KPCA. As with the dependence on resolution, this suggests that KPCA is (usefully) using information on small scales.

 \item Turning to cross-correlations with galaxy surveys, the advantages KPCA has over PCA/ICA largely vanish. For the auto-power spectrum, KPCA is significantly better than PCA in several cases, but in cross-correlation they are both similar, with less signal loss observed for PCA on large scales.
\end{itemize}
In conclusion, we find that the KPCA method has substantially different behaviour from the (linear) PCA and ICA methods when applied to the recovery of the 21cm auto-power spectrum. This makes it a useful alternative, e.g. for comparison studies used to verify the robustness of results to foreground removal. The KPCA method performs better than PCA and ICA in most (but not all) scenarios, specifically when the Sigmoid kernel is used. The kernel hyperparameters can be tuned using simulations to optimise performance, and the best choice of parameters will depend on the properties of the data (e.g. its resolution). The method is robust to masks and missing data, and does not require the data to be smoothed to a common resolution, as in fact it appears to benefit from the presence of small-scale information. KPCA performs almost equally as well as PCA or ICA when applied to the recovery of the galaxy-21cm cross-power spectrum however, further underlining the expected robustness of cross-correlation analyses.

\section*{Acknowledgements}

We are grateful to S.~Paul, P.~S.~Soares, M.~Spinelli, and E.~R.~Switzer for useful discussions. This project has received funding from the European Research Council (ERC) under the European Union’s Horizon 2020 research and innovation programme (grant agreement No.~948764). We acknowledge use of the following software: {\tt corner} \citep{corner}, {\tt HEALPix} \citep{Gorski:2004by}, {\tt matplotlib} \citep{matplotlib}, {\tt nbodykit} \citep{nbody}, {\tt numpy} \citep{numpy}, {\tt scikit-learn} \citep{scikit-learn}, and {\tt scipy} \citep{2020SciPy-NMeth}.

\section*{Data Availability}

The Python code used to produce the results in this paper is available from \url{https://github.com/philbull/FastBox}. The notebooks and simulated data used in this paper will be made available on reasonable request to the corresponding author.

\balance



\bibliographystyle{mnras}
\bibliography{kernelpca}





\bsp	
\label{lastpage}
\end{document}